\documentclass[12pt]{iopart}
\usepackage{iopams}
\usepackage{graphicx}
\input amssym.def \input amssym

\begin{document}


\title[]{Clifford group dipoles and the enactment of Weyl/Coxeter group $W(E_8)$ by entangling gates}

\author{Michel Planat
}

\address{Institut FEMTO-ST, CNRS, 32 Avenue de
l'Observatoire,\\ F-25044 Besan\c con, France (planat@femto-st.fr)}

 

\begin{abstract}

Peres/Mermin arguments about no-hidden variables in quantum mechanics are used for displaying a pair $(R,S)$ of entangling Clifford quantum gates, acting on two qubits. From them, a natural unitary realization of Coxeter/Weyl groups $W(D_5)$ and $W(F_4)$ emerges, which is also reflected into the splitting of the $n$-qubit Clifford group $\mathcal{C}_n$ into group dipoles $\mathcal{C}_n^{\pm}$. The union of the three-qubit real Clifford group $\mathcal{C}_3^+$ and the Toffoli gate ensures a orthogonal realization of the Weyl/Coxeter group $W(E_8)$, and of its relatives. Other concepts involved are complex reflection groups, $BN$ pairs, unitary group designs and entangled states of the $GHZ$, $W$ and chain families. 

\end{abstract}


\section{Introduction}

One important feature of quantum mechanics, not present in classical physics, is the possible non-commutativity of observables. Many peculiarities and paradoxes encountered in quantum mechanical measurements may be discussed in a discrete setting in terms of tensor products of Pauli spin matrices $\sigma_x$, $\sigma_y$ and $\sigma_z$, and the identity matrix $I$, which obey non-trivial commutation relations \cite{Pauligraphs}. In essence, the peculiarities carried by the Bell-Kochen-Specker theorem deal about the algebraic structure of eigenvalues/measurements, that contradicts that of the eigenstates \cite{Mermin93}.
In this note, the ingredients of this discussion are given a quantum computing setting.

We first exhibit a pair of two-qubit entangling gates $R$ (a braiding matrix) and $S$ (a non-braiding matrix), that encapsulate Mermin's discussion about quantum paradoxes. The octahedral geometry of the group $\left\langle R,S\right\rangle$ is investigated. Then, the natural decomposition of the symplectic group $\mbox{Sp}(2n,2)$ into its orthogonal subgroups \cite{Bannai99} is reflected in the splitting of the Clifford group $\mathcal{C}_n$ on $n$ qubits into group dipoles indexed by $R$ and $S$. The relevant Coxeter groups, BN pairs and unitary group designs rise up in the calculations.

Finally, it is shown that the real three-qubit group dipole, when complemented by the Toffoli gate, enact (up to isomorphism) the largest Weyl/Coxeter group $W(E_8)$ and its reflection subgroups. The various entangling gates spanning them are displayed.    

All group theoretical calculations are performed in Magma \cite{MAGMA}.

In the present paper, the term representation, or realization, of a group $G$ denotes a group homomorphism from $G$ to the general linear group $GL(n,\mathbb{C})$ of $n \times n$ matrices with complex entries. In Sec. 4, real representations, i.e. group homomorphisms from $G$ to $GL(n,\mathbb{R}$) will occur. The term {\it enactment} is used to contrast the new quantum gate presentation of $W(E_8)$, found in Sec. 4, with the standard realization, that makes use of the roots of the exceptional algebra $E_8$. We also coin the word {\it  Clifford group dipole}, in Sec. 3, for featuring the splitting of the Clifford group into its subgroups of positive/negative index. To our knowledge, this terminology was not used before. 

Appendices are devoted to the fundamental concepts underlying many aspects of the paper: complex reflection groups \cite{PlanatKibler08}, BN pairs \cite{PlanatSole08}, unitary designs \cite{Roy08} and the classification of entanglement \cite{Wootters00}.

\section{From Mermin's array to octahedral symmetry}

The basic pieces of the proof of Kochen-Specker theorem in a four-dimensional space are two triples of (mutually commuting and real) two-qubit observables
\begin{equation}
\left\{\sigma_x\otimes \sigma_x,\sigma_y\otimes \sigma_y,\sigma_z\otimes \sigma_z\right\}~\mbox{and}~\left\{\sigma_x\otimes \sigma_z,\sigma_z\otimes \sigma_x,\sigma_y\otimes \sigma_y\right\}.
\label{eq1}
\end{equation}
See \cite{Mermin93} for a full discussion based on the $3 \times 3$ Mermin's array and \cite{Pauligraphs} for the interpretation of the array as a specific  hyperplane of the generalized quadrangle of order two.  The joined eigenstates of the first triple of mutually commuting observables in (\ref{eq1}) are made explicit in \cite{Planat06}.   They may be casted as the rows of the orthogonal matrix $R$ as below
\begin{equation}
R=\frac{1}{\sqrt{2}}\left(\begin{array}{cccc} 1 & 0 & 0 & 1 \\0 & 1        & -1 & 0 \\ 0 & 1 & 1 & 0 \\-1 & 0 & 0 & 1\\ \end{array}\right),~~
\left(\begin{array}{ccc} + & + & - \\- &-        & -  \\ - & + &+  \\+ & - & + \\ \end{array}\right).
\end{equation}
Rows of the second matrix contain the sign of eigenvalues $\pm 1$, and the action of the transpose $R^{tr}$ of matrix $R$ on the computational basis leads to the entangled states shared by the triple, i.e. $R^{tr}\left|00\right\rangle=\frac{1}{\sqrt{2}}(\left|00\right\rangle+\left|11\right\rangle), \cdots$. Matrix $R$ is known as the Bell basis change matrix. It occurs in the braiding approach of quantum computing \cite{Kauf04,Werner09} and is also encountered in our recent papers \cite{PlanatJorrand08}-\cite{PlanatKibler08}.

The joined eigenstates of the second triple of mutually commuting observables in (\ref{eq1}) may be similarly casted as the rows of the entangling orthogonal matrix
\begin{equation}
S=\frac{1}{2}\left(\begin{array}{cccc} 1 & -1 & 1 & 1 \\1 & 1        & -1 & 1 \\1 & -1 & -1 & -1 \\1 & 1 & 1 & -1\\ \end{array}\right),~~
\left(\begin{array}{ccc} + & - & - \\- &+        & -  \\ - & - & +  \\+ & + & + \\ \end{array}\right).
\end{equation}
Thus the two matrices $R$ and $S$ capture, in a very compact and yet unnoticed form, the ingredients contained in the Mermin's proof of Kochen-Specker theorem. As any entangling matrix associated with local unitary transformations is universal for quantum computation, one can deduce that matrices $R$ and $S$, separately, are universal  (see Sec. 2.1 in \cite{Kauf04}). The $R$ matrix satisfies the Yang-Baxter equation (see Sec. 2 of \cite{Kauf04}), but the $S$ matrix does not. It has a different status, that is used for deriving the largest cristallographic group $W(E_8)$ in Sec. (\ref{sectionE8}). Both matrices are related by a relation involving the Hadamard matrix $H$ as
\begin{equation}
RS=H\otimes I~\mbox{with}~ H=\frac{1}{2}\left(\begin{array}{cc} 1 & 1 \\1         & -1\end{array}\right).
\end{equation}
Matrices $R$ and $S$ are {\it distinguished} members of the two-qubit Clifford group $\mathcal{C}_2$. Recall that the $n$-qubit Clifford group $\mathcal{C}_n$ is defined as the normalizer, in the unitary group $U(2^n,\mathbb{C})$, of the $n$-qubit Pauli group $\mathcal{P}_n$. The $n$-qubit Pauli group $\mathcal{P}_n$ is generated by the $n$-fold tensor products of ordinary Pauli spin matrices. Thus, by definition, the two-qubit Pauli group $\mathcal{P}_n$ maps to itself under the conjugation action of matrices/gates of $\mathcal{C}_n$. The distinctive character of $R$ and $S$ leads to a natural splitting of the Clifford group into two poles (see \cite{Jozsa08} for details about the Clifford group and \cite{PlanatSole08} for preliminary observations about its decomposition into BN-pairs \cite{Tits74}).

Beforehand, the finite group $G_{96}$ generated by $R$ and $S$ is of order $96$ and satisfies the group isomorphisms
\begin{equation}
G_{96}=\left\langle R,S\right\rangle \cong \mathcal{U}_{13} \cong \mathbb{Z}_4.\mathbb{S}_4,
\label{dot}
\end{equation}
where $\mathcal{U}_{13}$ is the complex reflection group No $13$ in the Shephard-Todd sequence \cite{Kane01} (see appendix A for a definition and \cite{PlanatKibler08} for a recent essay about the relevance of complex reflection groups for quantum information). From (\ref{dot}) it is clear that the group $G_{96}$ possesses a normal subgroup isomorphic to the cyclic group $\mathbb{Z}_4$. The dot product of groups means that the extension is not split and that its quotient by $\mathbb{Z}_4$  is  the symmetric group on four letters $\mathbb{S}_4$. The single qubit Clifford group $\mathcal{C}_1$ contains the reflection group $\mathcal{U}_{13}$ as a maximal subgroup of index two. The reflection group $\mathcal{U}_8$, No 8 in the Shephard-Todd sequence, is also a maximal subgroup of $\mathcal{C}_1$, has the same order, and is isomorphic to $\mathbb{Z}_4.\mathbb{S}_4$ as well. The latter is studied in \cite{Bannai99} in connection to the modular invariance property of association schemes and the related self-dual classical codes \cite{Nebe01}.

The smallest degree invariant of the Clifford group $\mathcal{C}_1$ is shared by its reflection subgroups $\mathcal{U}_8$ and $\mathcal{U}_{13}$ as 
\begin{equation}
\mathcal{W}:=\alpha^8+14\alpha^4\beta^4+\beta^8.
\label{inv3}
\end{equation}
It corresponds to the octahedral invariance. It was derived for the first time in 1913 Klein's lectures \cite{Klein56}. The octic invariant (\ref{inv3}) is obtained by exploiting the invariance under $SU(2)$ transformations of the center of faces of an octahedron with vertices located on the Riemann sphere. Then, seing the Riemann sphere as the usual Bloch sphere the variables $\alpha$ and $\beta$ may be interpreted as the amplitudes of a single qubit state. Invariant $\mathcal{W}$ may also be seen as the complete weight enumerator of the self dual code $e_8$ \cite{Nebe01}. 

The corresponding octic invariant of the (rank four) group $G_{96}$ is
\begin{equation}
W^{(2)}:=\Sigma_8+14 \Sigma_{4,4}+168 \Sigma_{2,2,2,2},
\label{inv3_2}
\end{equation}
in the notations of \cite{Nebe01}, i.e. $\Sigma_8=\sum_{i=1}^4 \alpha_i^8$, $\Sigma_{4,4}=\sum_{j>i} \alpha_i^4 \alpha_j^4$ and $\Sigma_{2,2,2,2}=\prod_{i=1}^4 \alpha_i^2$. Invariant $W^{(2)}$ represents the complete weight enumerator of the (genus two) code $e_8\otimes \mathbb{F}_4$ \cite{Nebe01}, and indeed generalizes Klein's invariant (\ref{inv3}).

 \section{Clifford group dipoles}
 
Let us call $\mathcal{P}_n$ the Pauli error group on $n$ qubits and $\mathcal{C}_n$ the corresponding Clifford group, i.e. the normalizer of $\mathcal{P}_n$ in the general $2^n$-dimensional complex unitary group $U(2^n,\mathbb{C})$. By definition, $\mathcal{P}_n$ is mapped to itself under the conjugation action $\mathcal{C}_n\mathcal{P}_n\mathcal{C}_n^{-1}$. The group $\mathcal{P}_n$ is normal in $\mathcal{C}_n$ and the map from $\mathcal{P}_n$ to the factor group $\mathcal{C}_n/\mathcal{P}_n$ is an element of the symplectic group $\mbox{Sp}(2n,2)$ \cite{Vourdas07}.

Then, the natural decomposition of $\mbox{Sp}(2n,2)$ into its orthogonal subgroups $\Omega^{\pm}(2n,2)$, studied in \cite{Bannai71} \footnote{The orthogonal group $\Omega^{\pm}(2n,2)$, of even dimension $2n$ ($n\ge 1$), is defined over the field $GF(2)$ in terms of two generating matrices. It is the kernel of the spinor map on the orthogonal group $SO^{\pm}(2n,2)$, and may also be defined as the derived subgroup of the general orthogonal group $O^{\pm}(2n,2)$. It is usually a perfect group.}, may be used for obtaining a straightforward decomposition of the Clifford group $\mathcal{C}_n$ into dipole subgroups $\mathcal{C}^{\pm}_n$
\begin{equation}
\mathcal{C}^{\pm}_n=E^{\pm}(2n+1).\Omega^{\pm}(2n,2).
\label{dip}
\end{equation}
in which $E^{\pm}(2n+1)$ are the extraspecial groups \cite{Nebe01}, of order $2^{2n+1}$, of the Pauli group $\mathcal{P}_n$. The splitting of the Clifford group into its subgroup dipoles is inspired by our recent proposal of decomposing the Clifford group into $BN$-pairs \cite{PlanatSole08}.

\subsection*{Single qubit group dipoles}
The single qubit Clifford group possesses a BN-pair (see the appendix B for the meaning of a BN-pair)
\begin{equation}
B \equiv C_1^+,~~ N\equiv C_1^-\cong \mbox{SL}(2,3),~~H_0\cong \mathbb{Z}_4~~\mbox{and}~~W\cong W(D_2)\cong \mathbb{Z}_2^2, 
\end{equation}
corresponding to the group dipoles
\begin{equation}
C_1^+= E^+(8),~~C_1^-\cong E^-(8).\Omega^-(2,2).
\end{equation}
The group dipole $\mathcal{C_1}\-=\mathcal{C_1}'$ (with $'$ meaning the derived subgroup) is nothing but the octahedral group
\begin{equation}
\mathcal{O}=\left\langle i\sigma_z,\frac{1}{2}\left(\begin{array}{cc} 1-i & i-1 \\ 1+i & 1+i  \\ \end{array}\right)
\right\rangle,
\label{octah}
\end{equation}
that is not of the reflection type, but isomorphic to the complex reflection group $\mathcal{U}_4$, No $4$ in the Shephard-Todd sequence. In virtue of its isomorphism to $SL(2,3)$, the group dipole $\mathcal{C}_1^-$ also corresponds to the smallest known unitary two-dimensional $2$-design (see \cite{Gross07,Roy08} and the appendix C for the definition).

\subsection*{Two-qubit group dipoles}

The relevant groups are the Clifford group $\mathcal{C}_2$ (order $92160$) and two subgroups: the {\it local} Clifford group $\mathcal{C}_2^{L} \supset S $ (order $4608$) and the Bell group $\mathcal{B}_2 \supset R$ (order $15360$), which may be generated as

\begin{eqnarray}
&\mathcal{C}_2=\left\langle \mathcal{C}_1 \otimes \mathcal{C}_1, \mbox{CZ}\right\rangle= \left\langle H\otimes H, H \otimes P, \mbox{CZ}\right\rangle,\nonumber\\
&\mathcal{C}_2^{L}=\left\langle \mathcal{C}_1 \otimes \mathcal{C}_1\right\rangle=\left\langle H\otimes H, H \otimes P, T\right\rangle,\nonumber \\
&\mathcal{B}_2=\left\langle H\otimes H, H \otimes P, R\right\rangle,\nonumber\\
\end{eqnarray}
with $\mbox{CZ}=\mbox{diag}(1,1,1,-1)$, $P$ is the $\pi/2$ phase gate and $T$ is the swap gate \cite{Jozsa08}. The importance of the {\it match} gate $R$ and of the {\it swap} gate $T$, in the context for the BN pair decomposition of the Clifford group, is an important issue of \cite{PlanatSole08}. The inclusion of matrices $R$ in $\mathcal{B}_2$ and $S$ in $\mathcal{C}_2^L$ is exclusive, i.e. $R\notin \mathcal{C}_2^L $ and $S\notin \mathcal{B}_2$. The pair $(R,S)$ of universal matrices  of the preceeding section reflects into pairs of subgroups of the Clifford group. In the sequel, one may see the positive group dipole $\mathcal{C}_2^+$ (which contains $S$ and not $R$) as indexed by $S$ and the negative group dipole $\mathcal{C}_2^-$ (which contains $R$ and not $S$) as indexed by $R$.

More precisely, one can construct a BN-pair of the two-qubit Clifford group \cite{PlanatSole08}
\begin{equation}
B\cong W(F_4),~N\equiv \mathcal{B}_2,~H_0\cong \mathbb{Z}_8~\mbox{and}~W\cong W(D_5),
\end{equation}
\noindent
in which $B$ is the Coxeter group of type \lq\lq$F_4$" (the symmetry group of the $24$-cell), $N$ is the Bell group, and the Weyl group W of the pair is isomorphic to the Coxeter group of type \lq\lq$D_5$."
The two-qubit Clifford group dipoles are
\begin{eqnarray}
&\mathcal{C}_2^+\cong E^+(32).\Omega^+(4,2)\cong E^+(32) \rtimes \mathbb{S}_3^2 \cong W(F_4),\nonumber \\
&\mathcal{C}_2^-=\mathcal{B}'_2\cong E^-(32).\Omega^-(4,2)\cong E^-(32).A_5,\nonumber  \\
\end{eqnarray}
\noindent

There exists an important maximal subgroup of the group dipole $\mathcal{C}_2^-$, that is isomorphic to the group $SL(2,5)$ and, as  the alternating group $A_5$, corresponds to the smallest known $2$-dimensional $5$-design \cite{Roy08} with generators

\footnotesize
\begin{equation}
\frac{1}{2}\left(\begin{array}{cccc} 1+i & 1-i & 0 & 0 \\-1-i & 1-i        & 0 & 0 \\ 0 & 0 & 1+i & 1-i \\0 & 0 & -1-i & 1-i\\ \end{array}\right),~~\frac{1}{2}\left(\begin{array}{cccc} 0 & 1+i & 1+i & 0 \\i-1 & 0        & 0 & 1-i \\i-1 & 0 & 0 & i-1 \\0 & -1-i & 1+i & 0\\ \end{array}\right).
\end{equation}
\normalsize

Their action on the computational base is either separable (for the l.h.s. generator) or gives rise to a Bell basis with a phase factor (for the r.h.s. generator).

\subsection*{Three-qubit and higher-order group dipoles}
The three-qubit Clifford group dipoles are
\begin{eqnarray}
&\mathcal{C}_3^+\cong E^+(128).\Omega^+(6,2)\cong E^+(128) .A_8,\nonumber \\
&\mathcal{C}_3^-=\mathcal{B}'_3\cong E^-(128).\Omega^-(6,2)\cong E^-(128).W'(E_6),\nonumber \\
\end{eqnarray}
\noindent
in which the simple groups $A_8$ (the eight-letter alternating group) and $W'(E_6)$ (the derived subgroup of the Coxeter group of type \lq\lq $E_6$" ) are of order $20160$ and $25920$, respectively. More generally, higher order group dipoles involve the orthogonal groups $\Omega^{\pm}(2n,2)$, that identify to $D_n(2)$ and $2D_n(2)$ (twisted) Chevalley groups.

The two constitutive real entangling gate $S$ ant swap gate $T$ may be used for generating the group dipole $\mathcal{C}_3^+$, of order $2~580~480$, as follows
\begin{equation}
\mathcal{C}_3^+=\left\langle \sigma_x \otimes S, S\otimes \sigma_x, \sigma_x \otimes T, T\otimes \sigma_x\right\rangle.
\end{equation}
See also \cite{Nebe01} for the relation between real Clifford groups and Barnes-Wall lattices.

Among the subgroups of $\mathcal{C}_3^+$ having relevance to unitary group designs, there are five subgroups, isomorphic to the complex reflection group $U_{24}$ (of order $336$). As their central quotient, isomorphic to the simple subgroup $G_{168}=PSL(2,7)$, they define three-dimensional $2$-designs. The smallest known design with these parameters is of order $72$, isomorphic to $\mathbb{Z}_3^2 \rtimes E^-(8)$ \cite{Roy08}, and is also a subgroup of $\mathcal{C}_3^+$. 

The simple group $G_{168}$ is intimately related to tripartite entanglement (see the appendix D for a reminder about the measures of entanglement). In the representation
 
\footnotesize
\begin{equation}
G_{168}=\left\langle a=\sigma_0\otimes \mbox{CZ},~~
b=\frac{1}{2}\left(\begin{array}{cccccccc} 1&1&0&0 &0&0&-1&1 \\1&1&0&0 &0&0&1&-1\\0&0&1&-1 &-1&-1&0&0\\0&0&-1&1 &-1&-1&0&0\\
-1&1&0&0 &0&0&1&1 \\-1&1&0&0 &0&0&-1&-1\\0&0&-1&-1 &1&-1&0&0\\0&0&1&1 &1&-1&0&0\\ \end{array}\right)\right\rangle ,
\label{matrixb}
\end{equation}
\normalsize
with orthogonal matrices $a$ and $b$ one observes that the action of $b^{tr}$ on the computational basis creates a $8$-dim basis of GHZ states such as the entangled state
\begin{equation}
b^{\tr}\left|000\right\rangle=\frac{1}{2}(\left|000\right\rangle+\left|001\right\rangle-\left|110\right\rangle+\left|111\right\rangle).
\end{equation}
Using concepts recalled at the appendix D, one easily obtains the residual tangle $\tau^{(3)}=1$ and the vanishing of all bipartite tangles, that are properties specific to quantum states of the GHZ family. With further insight \footnote{The mutually commuting sets encoding the matrices $b$ in (\ref{matrixb}), $c$ and $d$ in (\ref{matrixc}) were discovered by Peter Levay, see also \cite{Levay08}.}, it is found that rows of matrix $b$ correspond the joined eigenstates of the following set of mutually commuting operators 
\begin{eqnarray}
&I\otimes \sigma_z\otimes\sigma_x,~\sigma_x\otimes \sigma_x \otimes \sigma_z,~\sigma_x\otimes \sigma_y \otimes \sigma_y,~\sigma_z\otimes I \otimes \sigma_x, \nonumber \\
&\sigma_z\otimes \sigma_z \otimes I,~\sigma_y\otimes \sigma_x \otimes \sigma_y,~\sigma_y\otimes \sigma_y \otimes \sigma_z.
\label{common1}
\end{eqnarray}

The three-qubit representation of the simple group $PSL(2,7)$ in (\ref{matrixb}) is indeed quite different from the Hurwitz presentation with generators and relations $\left\langle x,y|x^2=y^2=(xy)^7=[x,y]^4=1\right\rangle$ (in which $[x,y]$ means the group commutator of elements $x$ and $y$) \cite{Conder90}. The presentation with generators and relations one obtains for $G_{168}$ is $a^2=b^4=(b a^{-1})^7=(b^{-2}a)^2=1$.

\section{Enactment the Weyl/Coxeter group $W(E_8)$ and its relatives}
\label{sectionE8}

The finite Coxeter/Weyl group of the largest cardinality is $W(E_8)\cong \mathbb{Z}_2.O^+(8,2)$, of order $696~729~600$, in which $O^+(8,2)$ is the general eight-dimensional orthogonal group over the field $GF(2)$ and $O^+(8,2)'=\Omega^+(8,2)$. It may be realized as the complex reflection group $\mathcal{U}_{37}$ ($2^{120}$ reflections), the last group in the Shephard-Todd sequence. Group $W(E_8)$ plays a unifying role in physics \cite{Koca01,Lisi07}, being the symmetry group of the largest exceptional root system, that of the simple Lie group $E_8$.

Until now, we have crossed many important subgroups of the Clifford group that are isomorphic to complex reflection groups and one may legitimately ask whether it is a mere coincidence, or if there is a deeper mechanism relating quantum computing and all the finite reflection groups. Remarkably, $W(E_8)$ can be generated by adjoining to the real Clifford group $\mathcal{C}_3^+$ the (non-Clifford) Toffoli gate $C^2\mbox{NOT}=\mbox{TOF}$, which applies a $\mbox{NOT}$ operation to the (target) third qubit only if the  two first (control) qubits are set to $\left|1\right\rangle $. Thus,
\begin{equation}
\left\langle \mathcal{C}_3^+,\mbox{TOF}\right\rangle =\left\langle I\otimes S,S \otimes I, \mbox{TOF}\right\rangle\cong W(E_8).
\label{Ent}
\end{equation}
Conversely, it is easy to recognize $\mathcal{C}_3^+$ as isomorphic to the second largest maximal subgroup of $W'(E_8)\cong O^+(8,2)$.

The Toffoli gate is a well known universal and reversible logic gate for classical computing. Since any reversible gate may be implemented on a quantum computer, it also serves as a quantum gate. The union of Toffoli and Hadamard gates is universal for quantum computation \cite{Shi02}.

Let us rewrite (\ref{Ent}) as   

\footnotesize
\begin{equation}
\left\langle b, \tilde{b}=\frac{1}{2}\left(\begin{array}{cccccccc} 1&-1&0&0 &0&0&1&1 \\1&1&0&0 &0&0&-1&1\\0&0&1&1 &-1&1&0&0\\0&0&-1&1 &-1&-1&0&0\\
0&0&1&1 &1&-1&0&0 \\0&0&-1&1 &1&1&0&0\\-1&1&0&0 &0&0&1&1\\-1&-1&0&0 &0&0&-1&1\\ \end{array}\right), \mbox{TOF} \right\rangle \cong W(E_8),
\end{equation}
\normalsize
where $\mathcal{C}_3^+=\langle b,\tilde{b}\rangle$ is an alternative way to generate the $3$-qubit real Clifford group with generators of the GHZ-type. One easily observes that $b$ and $\tilde{b}$ only differs from a reordering of the rows and thus correspond to the same set of eigenstates, already displayed in (\ref{common1}).

Let us now turn to another set of GHZ-type gates for generating the group $W(E_8)$
\footnote{One can check that matrix $c$ has rows encoding states shared by the following set of seven mutually commuting operators
\begin{eqnarray}
&\sigma_z \otimes \sigma_z\otimes\sigma_z,~I\otimes \sigma_y \otimes \sigma_y,~\sigma_y\otimes I\otimes \sigma_y,~\sigma_y\otimes \sigma_y \otimes I, \nonumber \\
&\sigma_z\otimes \sigma_x \otimes \sigma_x,~\sigma_x\otimes \sigma_z \otimes \sigma_x,~\sigma_x\otimes \sigma_x \otimes \sigma_z,
\label{common2}
\end{eqnarray}
and that matrix $d$ has rows encoding the shared eigenspace of the triple of mutually commuting operators
\begin{equation}
\sigma_y\otimes \sigma_z \otimes \sigma_y,~\sigma_y\otimes \sigma_y \otimes \sigma_x,~I\otimes \sigma_x \otimes \sigma_z.
\label{common3}
\end{equation}
}

\footnotesize
\begin{eqnarray}
&c=\frac{1}{2}\left(\begin{array}{cccccccc} 1&0&0&-1 &0&1&1&0 \\0&1&-1&0 &1&0&0&1\\0&1&1&0 &1&0&0&-1\\1&0&0&1 &0&1&-1&0\\
0&-1&1&0 &1&0&0&1 \\-1&0&0&1 &0&1&1&0\\1&0&0&1 &0&-1&1&0\\0&1&1&0 &-1&0&0&1\\ \end{array}\right),\nonumber \\
&d=\frac{1}{2}\left(\begin{array}{cccccccc} 1&0&-1&0 &0&-1&0&-1 \\0&1&0&1 &1&0&-1&0\\1&0&1&0 &0&-1&0&1\\0&-1&0&1 &-1&0&-1&0\\
0&-1&0&1 &1&0&1&0 \\1&0&1&0 &0&1&0&-1\\0&1&0&1 &-1&0&1&0\\1&0&-1&0 &0&1&0&1\\ \end{array}\right)\nonumber \\
\label{matrixc}
\end{eqnarray}
\normalsize
that are such that
\begin{equation}
\left\langle c,d, \mbox{TOF}\right\rangle \cong W(E_8)~,\mathcal{C}_3^+=\left\langle c,d,\sigma_0 \otimes \mbox{CZ} \right\rangle~\mbox{and}~ \left\langle c,d\right\rangle \cong \mathcal{C}_2^-. 
\end{equation}
The second largest complex reflection group $\mathcal{U}_{36}\equiv W(E_7)$, of order $2~903~040$ with $2^{63}$ reflections, may be generated as
\begin{equation}
\left\langle b,c, \mbox{TOF}\right\rangle \cong W(E_7)~\mbox{where}~ \left\langle b,c\right\rangle \cong \mathbb{Z}_2^4 \rtimes \mathbb{D}_4 ~\mbox{and}~\langle \tilde{b},c\rangle \cong \mathcal{C}_2^-, 
\end{equation}
where $D_4$ is the dihedral group of order $8$.
Still another way to realize/enact $W(E_8)$ in a unitary way is to complement the $3$-qubit representation of $SL(2,5)$ with the Toffoli gate as follows
\begin{equation}
\langle \tilde{b},d, \mbox{TOF}\rangle\cong W(E_8)~\mbox{with}~\mathcal{C}_3^+=\langle \tilde{b},d,\sigma_0 \otimes \mbox{CZ} \rangle~\mbox{and}~ \langle \tilde{b},d\rangle \cong SL(2,5).
\label{SL25}
\end{equation}
Indeed, the unitary realization of $W(E_8)$ with quantum gates of the GHZ type is much different from the Weyl group one gets from the Lie algebra of $E_8$. Recall that the gears of this new representation of $W(E_8)$, that induce the tripartite entanglement, are simply the real bipartite entangling matrix $X$ and the Toffoli gate, that we used in (\ref{Ent}). The Pandora's box of cristallographic groups only arises from these two conclusive players.

\subsection*{Enacting the Weyl/Coxeter group $W(E_6)$}

The Weyl/Coxeter group $W(E_6)\cong O^-(6,2)$, of order $51840$ with $2^{36}$ reflections, is an important subgroup of $W(E_7)$, being the symmetry group of a {\it smooth cubic surface} embedded in the three-dimensional complex projective space $\mathbb{P}^3(\mathbb{C})$. The surface contains a maximum of $27$ lines in general position and $45$ sets of tritangent planes. The group of permutations of the $27$ lines is $W(E_6)$, the stabilizer of a line is $W(D_5)$ (observe that $|W(E_6)|/|W(D_5)|=27$) and the stabilizer of a tritangent plane is $W(F_4)$ \cite{Hunt00}. Thus, the $\mbox{BN}$-pairs, and the Clifford group dipoles described at the previous section,are reflected into the geometry of such a cubic surface.

But $W(E_6)$ is not a subgroup of the $3$-qubit Clifford group, further gates have to be added to display it. One among many unitary realizations of $W(E_6)$ is 

$$\left\langle e,f, \mbox{TOF}\right\rangle \cong W(E_8)~\mbox{and}~ \left\langle e,f\right\rangle \cong W(E_6),\nonumber$$

\footnotesize
\begin{eqnarray}
&\mbox{where}~~e=\frac{1}{2}\left(\begin{array}{cccccccc} 0&0&0&-1 &-1&-1&0&-1 \\-1&-1&-1&0 &0&0&-1&0\\0&0&0&1 &1&-1&0&-1\\-1&-1&1&0 &0&0&1&0\\
-1&1&1&0 &0&0&-1&0 \\0&0&0&-1 &1&-1&0&1\\1&-1&1&0 &0&0&-1&0\\0&0&0&-1 &1&1&0&-1\\ \end{array}\right),\nonumber \\
&\mbox{and}~~f=\frac{1}{4}\left(\begin{array}{cccccccc} -1&-1&1&-1 &-1&1&3&-1 \\1&1&-1&1 &1&3&1&1\\1&1&-1&1 &1&-1&1&-3\\3&-1&1&-1 &-1&1&-1&-1\\
-1&-1&-3&-1 &-1&1&-1&-1 \\-1&3&1&-1 &-1&1&-1&-1\\-1&-1&1&3 &-1&1&-1&-1\\-1&-1&1&-1 &3&1&-1&-1\\ \end{array}\right).\nonumber \\
\label{WE6}
\end{eqnarray}

\normalsize
Similarly to the Toffoli gate, gates $e$ and $f$ do not belong to the real Clifford group $\mathcal{C}^+_3$. If one complements the unitary representation of $\mathcal{C}^+_3$, or the one of $W(E_7)$, by gate $e$, or  by gate $f$, or by both gates $e$ and $f$, one obtains a representation of $W'(E_8)$, of cardinality $|W(E_8)|/2$. Another useful expression is $\langle b,e\rangle \cong W(E_7)$.

The entanglement involved in the matrices is peculiar. As shown in the example provided at the appendix D, the entanglement for the states arising from matrix $e$ is a linear chain $A-B-C$, and a similar calculation for the states arising from the matrix $f$ shows that the entanglement of the $W$ type.

\section{Discussion}

I discussed a relationship between Mermin's approach of Kochen-Specker theorem and quantum computation. I introduced a bipolar decomposition of the Clifford group, attached to error correction, and made explicit the corresponding generating gates. A new orthogonal realization of Weyl/Coxeter group $W(E_8)$ based on quantum gates has been uncovered. It opens up new vistas for quantum computing by providing optimal sets of gates with a clear group theoretical structure, such as BN-pairs and designs, that may serve for specific purposes. It also adds an alternative to the complex reflection groups of the Shephard-Todd list that often serve as the background of essays about the unification of physics \cite{Lisi07}. The peculiar role of the $2$-dimensional $5$-design $SL(2,5)$ in the cosmological context \cite{Kramer05} and its relation to $W(E_8$) is intriguing.  All types of three-qubit distributed entanglement arise in our unitary realization of $W(E_8)$ and of its reflection subgroups. See also \cite{Levay08} and references therein for the mathematical analogy between some stringy black hole solutions, quantum entanglement, finite geometries, and \cite{PlanatCPT} for further developments of the present paper.

\subsection*{Appendix A: Complex reflection groups}
Basically, reflections are linear transformations that leave invariant a hyperplane of a vector space, while sending vectors orthogonal to the hyperplanes to their negatives. For an Euclidean vector space $\mathcal{E}$, finite groups of reflections possess a {\it Coxeter group} structure, i.e. a a presentation in terms of a finite set of involutions with specific relations. There is a formal similarity between quantum errors $g$ of the Pauli group $\mathcal{P}_n$ and reflections $s_{\alpha}$ acting on the Euclidean space $\mathcal{E}$, and between the Clifford group action on $\mathcal{P}_n$ and the action of the orthogonal group $O(\mathcal{E})$ on $\mathcal{E}$.
 
A unitary element of the Clifford group maps $\mathcal{P}_n$ to itself 
 $$\forall g \in \mathcal{P}_n ~\mbox{and}~\mathcal{C} \in \mathcal{C}_n ,~\mathcal{C} g \mathcal{C}^{-1}=g'\in \mathcal{P}_n, $$
\noindent
and the orthogonal group $O(\mathcal{E})$ map reflections to reflections
 $$\forall s_{\alpha} \in O(\mathcal{E})~  \mbox{and}~ t \in O(\mathcal{E}),~t s_{\alpha}t^{-1}=s_{t_{\alpha}},  $$
\noindent
in which $\alpha$ denotes the index of a hyperplane of $\mathcal{E}$ and $t(\alpha)$ the index of the hyperplane mapped by the action of $O(\mathcal{E})$.
 
Euclidean reflection groups may be generalized to pseudo-reflection groups by replacing $\mathcal{E}$ by a vector space over the complex field $\mathbb{C}$. Finite irreducible unitary reflection groups are classified: They include the (real) Coxeter groups [usually denoted $W(X_i)$ for the Coxeter type $X_i$], three infinite families $\mathbb{Z}_m=\mathbb{Z}/m\mathbb{Z}$, the symmetric groups $\mathbb{S}_n$, the imprimitive reflection groups $G(m,p,n)=A(m,p,n)\rtimes \mathbb{S}_n$ (that are semi-direct products of a specific group of diagonal matrices with $\mathbb{S}_n$), and $34$ exceptional (Shephard-Todd) groups $\mathcal{U}_n$ \cite{Kane01}. The largest one is $\mathcal{U}_{37} \cong W(E_8)$, of cardinality $696~729~600$ with $2^{120}$ reflections.

Many of the self-dual codes so far derived rely on the well developed invariant theory of reflection groups and its relevance to Clifford groups \cite{Nebe01,PlanatKibler08}.

\subsection*{Appendix B: BN-pairs} 

Let consider a finite group $G$, and two subgroups $B$ and $N$ of $G$ generating $G$, $H_0=B \cap N$ a normal subgroup of $N$ and the quotient group  $W=N/H_0$ generated by a set $S\subset W$ of involutions. A group $G$ is said to have a $BN$-pair iff it is generated as above  and two extra relations (i) and (ii) are satisfied by the double cosets
 $$(\mbox{i})~~\mbox{For}~\mbox{any}~s \in S~\mbox{and}~w \in W,~ sBw \subseteq (BwB)\cup(BswB),$$
 $$(\mbox{ii})~~\mbox{For}~\mbox{any}~s \in S, sBs\nsubseteq B.$$
\noindent
The pair $(W,S)$ arising from a BN-pair is a Coxeter system.

One can form BN-pairs of the Clifford group \cite{PlanatSole08}.

\subsection*{Appendix C: Unitary designs} 

A unitary design is a set of unitary matrices that {\it simulates} the entire unitary group. It is a variation of spherical-$t$ designs and of Grassmannian $t$-designs \cite{Gross07,Roy08}.

In a unitary $t$-design, the integral $\int _{U(d)}U^{\otimes t}\otimes (U^*)^{ \otimes t} dU$ over all $d$-dimensional unitary matrices is identical to its restriction to a discrete subset  $X$.

For a finite set $X\subset U(d)$ of unitary matrices
%
$$\frac{1}{|X|^2} \sum_{U,V \subset X}\left|\mbox{tr}(U^*V)\right|^{2t}\ge \int_{U(d)}\left|\tr(U)\right|^{2t} dU,$$ 
%
with equality if and only if $X$ is a $t$-design.

Many unitary group designs were constructed as the images of unitary representations of finite groups \cite{Gross07,Roy08} using the following theorem:

Let $G$ be a finite group and $\rho:G\rightarrow U(d)$ a representation with character $\kappa$. Then $X=\left\{\rho(g):g \in G\right\}$ is a unitary $t$-design iff
$$\frac{1}{G}\sum_{g \in G}\left|\kappa(G)\right|^{2t}=\int_{U(d)}\left|\tr(U)\right|^{2t} dU. $$
The right hand side of the above equation is the moment of order $2t$ of the trace of a random $d$-dimensional unitary matrix. There exists a combinatorial interpretation as the number of permutations of length $t$ with no increasing subsequence of length greater than $d$. If $d\ge t$, then the r.h.s. is $t!$.
 
Many efficient designs may be harversted using the known character tables of finite groups. But several optimal (i.e. small size) Clifford designs could only be obtained from subgroups of the symplectic group $Sp(2n,q)$ acting transitively on a punctured vector space. The latter are closely related to the Clifford group dipoles investigated in this paper.

\subsection*{Appendix D: Measures of entanglement}
\label{entang}

The resources needed to create a given entangled state may be quantified, and one can define invariants for discriminating the type of entanglement.

For a pair of quantum systems $A$ and $B$ in a pure state of density matrix $\left|\psi \right\rangle \left\langle \psi \right|$, the {\it entanglement of formation} is defined as the entropy of either of the two subsystems $A$ and $B$
%
$$E(\psi)=-\mbox{tr}(\rho_A \log_2 \rho_A)=-\mbox{tr}(\rho_B \log_2 \rho_B),$$
%
where $\rho_A$ and $\rho_B$ are partial traces of $\rho$ over subsystems $B$ and $A$, respectively.
The measure is made explicit by defining the spin-flipped density matrix \cite{Wootters00} 
%
$$\tilde{\rho}=(\sigma_y \otimes \sigma_y)\rho^{\ast}(\sigma_y \otimes \sigma_y), $$
%
and the concurrence $C(\psi)=|\langle \psi |\tilde{\psi}\rangle|$ between the original and flipped state $\tilde{\psi}=\sigma_y \left|\psi^{\ast}\right\rangle$. As both $\rho$ and $\tilde{\rho}$ are positive operators, the product $\rho\tilde{\rho}$ also has only real and non-negative eigenvalues $\lambda_i$ (ordered in decreasing order) and the concurrence reads 
%
$$C(\rho)=\mbox{max}\left\{0,\sqrt{\lambda_1}-\sqrt{\lambda_2}-\sqrt{\lambda_3}-\sqrt{\lambda_4}\right\}. $$
%

For a two-qubit state  $\left| \psi\ \right\rangle=\alpha \left|00\ \right\rangle + \beta\left|01\ \right\rangle+ \gamma\left|10\ \right\rangle+ \delta\left|11\ \right\rangle$, the concurrence is $C=2\left|\alpha\delta-\beta\gamma\right|$, and thus satisfies the relation $0\le C \le 1$, with $C=0$ for a separable state and $C=1$ for a maximally entangled state.

The entanglement of a triple of quantum systems $A$, $B$ and $C$ in a pure state may be conveniently described by tracing out over partial subsystems $AB$, $BC$, and $AC$. In this generalized context, one introduces the {\it tangle} $\tau=C^2$. Tangles attached to the bipartite subsystems above satisfy the inequality

%
$$\tau_{AB}+\tau_{AC}\le 4 \mbox{det}\rho_A \equiv \tau_{A(BC)}. $$
%
The right hand side is interpreted as the amount of entanglement shared by the single qubit $A$ with the pair $BC$, in comparison with the amounts of entanglement shared with qubits $B$ and $C$ taken individually. It is remarkable that, for any value of the tangles satisfying this inequality, one can find a quantum state consistent with those values \cite{Wootters00}.

It has been shown that an arbitrary three-qubit state $\left|\psi\right\rangle$ can be entangled in essentially two inequivalent ways, belonging to the GHZ-class: $\left|\mbox{GHZ}\right\rangle=\frac{1}{\sqrt{2}}(\left|000\right\rangle+\left|111\right\rangle)$ or to the W-class: $\left|\mbox{W}\right\rangle=\frac{1}{\sqrt{3}}(\left|001\right\rangle)+\left|010\right\rangle+\left|100\right\rangle)$, according whether $\psi$ can be converted to the state $\left|\mbox{GHZ}\right\rangle$ or to the state $\left|\mbox{W}\right\rangle$, by stochastic local operations and classical communication (SLOCC) \cite{Dur00}. The relevant class is determined by computing the bipartite tangles of the reduced subsystems. If they vanish, then the subsystems are separable and
$\left|\psi\right\rangle$ belongs to the GHZ-class, meaning that all the entanglement is destroyed by tracing over one subsystem. If none of the bipartite tangles vanish, then $\left|\psi\right\rangle$ belongs to the W-class, meaning that it maximally retains bipartite entanglement after tracing over one subsystem.

Further discrimination of the entanglement type of a general $3$-qubit state
%
$$\left|\psi\right\rangle=\sum_{a,b,c=0,1}\psi_{abc}\left|abc\right\rangle, $$
%
can be obtained by calculating the SLOCC invariant three-tangle \cite{Wootters00}
%
$$\tau^{(3)}=4\left|d_1-2d_2+4d_3\right|,$$
$$ d_1=\psi_{000}^2\psi_{111}^2+\psi_{001}^2\psi_{110}^2+\psi_{010}^2\psi_{101}^2+\psi_{100}^2\psi_{011}^2,$$
$$d_2=\psi_{000}\psi_{111}(\psi_{011}\psi_{100}+\psi_{101}\psi_{010}+\psi_{110}\psi_{001})$$
$$+\psi_{011}\psi_{100}(\psi_{101}\psi_{010}+\psi_{110}\psi_{001})+\psi_{101}\psi_{010}\psi_{110}\psi_{001},$$
$$d_3=\psi_{000}\psi_{110}\psi_{101}\psi_{011}+\psi_{111}\psi_{001}\psi_{010}\psi_{100}. $$
%
For the GHZ state the $3$-tangle becomes maximal: $\tau^{(3)}=1$ and it vanishes for any factorized state. It also vanishes for states of the $W$-class.
The $3$-tangle may be interpreted as the {\it residual tangle}
%
$$\tau^{(3)}=\tau_{A(BC)}-(\tau_{AB}+\tau_{AC}), $$
%
i.e., the amount of entanglement between subsystems $A$ and $BC$ that cannot be accounted for by the entanglements of $A$ with $B$ and $C$ separately. It is of course independent on which qubit one takes as the reference of the construction. The GHZ state is a true tripartite entangled state so that no amount of entanglement is in the bipartite subsystems, as a result the residual entanglement is maximal. In contrast, for the states of the W-class the entanglement is of a pure bipartite type and $\tau^{(3)}=0$. Mixtures of GHZ and W states are studied in \cite{Lohmayer06}, where it is shown that while the amounts of inequivalent entanglement types strictly add up for pure states, the {\it monogamy} is in general lifted for mixed states because the entanglement can arise from different types of locally inequivalent quantum correlations.

Apart from pure tripartite entanglement (the $GHZ$ states) and equally distributed bipartite entanglement (the $W$ states), one can obtain a linear chain configuration of entanglement of the type $A-B-C$, where the two parties $(A,B)$  and $(A,C)$ are both entangled, but the parties $(A,C)$ are not. One example is the state

%
$$\left|\psi\right\rangle= \frac{1}{2}(\left|011\right\rangle+\left|100\right\rangle+\left|101\right\rangle+\left|111\right\rangle),$$
%
which arises in the context in the unitary realization of the group $W(E_6)$ in (\ref{WE6}).
Using the relations above one gets $\tau^{(3)}=1/4$ (i.e. a non equally distributed entanglement) and the tracing over two qubits may be calculated as 
\scriptsize
$$\rho_{BC}=\frac{1}{4}\left(\begin{array}{cccc} 1 & 1 & 0 & 1 \\1 & 1        & 0 & 1 \\ 0 & 0 & 0 & 0 \\1 & 1 & 0 & 2\\ \end{array}\right),~
\rho_{AB}=\frac{1}{4}\left(\begin{array}{cccc} 0 & 0 & 0 & 0 \\0 & 1        & 1 & 1 \\ 0 & 1 & 2 & 1 \\0 & 1 & 1 & 1\\ \end{array}\right),~
\rho_{AC}=\frac{1}{4}\left(\begin{array}{cccc} 0 & 0 & 0 & 0 \\0 & 1        & 0 & 1 \\ 0 & 0 & 1 & 1 \\0 & 1 & 1 & 2\\ \end{array}\right).~$$
%
\normalsize

The sets of eigenvalues for the first two matrices $\rho_{BC}$ and $\rho_{AB}$ are
 $\left\{\frac{1}{16}(3+2\sqrt{2}), \frac{1}{16}(3-2\sqrt{2}),0,0\right\}$ so that the corresponding concurrence for pairs $(B,C)$ and $(A,B)$ is strictly positive. In contrast, the set of square eigenvalues for the matrix $\rho_{AC}$ is $\left\{\frac{1}{16},\frac{1}{16} ,0,0\right\}$ and the corresponding concurrence  for the pair $(A,C)$ vanishes.
  
\section*{Acknowledgements}
The author thanks Peter Levay for finding the mutually commuting operators occuring in the three-qubit entangled matrices, and Maurice Kibler for his comments on the manuscript.


\section*{Bibliography}

\vspace*{.0cm} \noindent
\vspace*{-.1cm}
\begin{thebibliography}{10}

\bibitem{Pauligraphs}
Planat M and Saniga M 2008 On the Pauli graphs of $N$-qudits {\it Quant. Inf. Comp.} {\bf 8} 127.

\bibitem{Mermin93}
Mermin N~D 1993 Hidden variables and the two theorems of John Bell {\it Rev. Mod. Phys.} {\bf 65} 803.

\bibitem{Bannai99}
Bannai E 1999 Modular invariance property of association schemes, type II codes over finite rings and finite abelian groups and reminiscences of François Jaeger (a survey) { \it Ann. Institut Fourier} {\bf 49} 763.

\bibitem{MAGMA}Bosma W, Cannon J and  Playoust C 1997 The Magma algebra system {\it J. Symbolic Comput.} {\bf 24} 235.


\bibitem{Planat06}
Planat M, Saniga M and Kibler M~R 2006 Quantum Entanglement and Projective Ring Geometry {\it SIGMA} {\bf 2} Paper 066.

\bibitem{Kauf04}
Kauffman L~H and Lomonaco S~J 2004 Braiding operators are universal quantum gates {\it New J. Phys.} {\bf 6} 134. 

\bibitem{Werner09}
Ahlbrecht A, Georgiev L~S and Werner R F  2009 Implementation of Clifford gates in the Ising-anyon topological quantum computer {\it Phys Rev A} {\bf 79} 032311.

\bibitem{PlanatJorrand08}
Planat M and Jorrand P 2008 On group theory for quantum gates and quantum coherence {\it J. Phys. A: Math. Theor.} {\bf 41} 182001.

\bibitem{PlanatSole08}
Planat M and Sol\'e P 2008 Clifford groups of quantum gates, $BN$-pairs and smooth cubic surfaces {\it J. Phys. A: Math. Theor.} {\bf 42} 042003.

\bibitem{Tits74}
Tits J 1974 Buildings of Spherical Type and Finite BN-Pairs {\it Lect. Notes in Math.} {\bf 386} (Springer: Berlin).

\bibitem{PlanatKibler08}
Planat M and Kibler M 2008 Unitary reflection groups for quantum fault tolerance {\it J. Comp. Theor. Nanosci.} (to appear),  Preprint 0807.3650 [quant-ph].

\bibitem{Jozsa08}
Jozsa R and Miyake A 2008 Matchgates and classical simulation of quantum circuits {\it Proc. R. Soc.} {\bf 464} 3089.


\bibitem{Kane01}
Kane R 2001 {\it Reflection groups and invariant theory} (Berlin: Springer).

\bibitem{Nebe01}
Nebe G, Rains E~M and Sloane N~J~A  2001 The Invariants of the Clifford Groups
  {\it Designs, Codes and Cryptography} {\bf 24} 99.

\bibitem{Klein56}
Klein F 1956 {\it Lectures on the icosahedron and the solutions of equations of the fifth degree} (Dover, Ney York).

\bibitem{Vourdas07}
Vourdas A. 2007 Quantum systems with finite Hilbert space: Galois fields in quantum mechanics {\it J. Phys. A: Math. Theor.} {\bf 40} R285.

\bibitem{Bannai71}
Bannai E 1971 On some subgroups of the group $Sp(2n,2)$.{\it Proc. Japan Acad.} {\bf 47} 769.

\bibitem{Gross07}
Gross D, Audenaert and Eisert J 2007 Evenly distributed unitaries: on the structure of unitary designs {\it J. Math. Phys.} {\bf 48} 052104.

\bibitem{Roy08}
Roy A and Scott A J 2008  Unitary designs and codes {\it Des Codes Cryptogr} {\bf 53} 13. 

\bibitem{Wootters00}
Coffman V, Kundu J and Wootters W~K 2000 Distributed entanglement {\it Phys. Rev. A} {\bf 61} 052306.

\bibitem{Dur00}
D\"{u}r W, Vidal G and Cirac J~J 2000 Three qubits can be entangled in two inequivalent ways {Phys. Rev. A} {\bf 62} 062314.

\bibitem{Lohmayer06}
Lohmayer R, Osterloh A, Siewert J and Uhlman A 2006 Entangled Three-Qubit States without Concurrence and Three-Tangle {\it Phys. Rev. Lett.} {\bf 97} 260502.

\bibitem{Conder90}
Conder M 1990 Hurwitz groups: a brief survey {\it Bull. Am. Math. Soc.} {\bf 23} 359.

\bibitem{Shi02}
Shi S 2002 Both Toffoli and Controlled-NOT need little help to universal quantum computing {\it Quant. Inf. Comp.} {\bf 3} 84.

\bibitem{Hunt00}
Hunt B 2000 {\it The geometry of some special arithmetic quotients} (Springer, Berlin).

\bibitem{Koca01}
Koca M, Koc R and Al-Barwani M 2001 Noncrystallographic Coxeter group H4 in E8 {\it J. Phys. A: Math. Gen.} {\bf 34} 11201.

\bibitem{Lisi07}
Lisi G 2007 An eceptionnally simple theory of everything. Preprint 0711.0770 [hep-th].

\bibitem{Kramer05}
Kramer P 2005 An invariant operator due to F Klein quantizes H Poincaré's dodecahedral $3$-manifold {\it J. Phys. A: Math. Gen.} {\bf 38} 3517.

\bibitem{Levay08}
Levay P, Saniga M and Vrana P 2008 Three-qubit operators, the split Cayley hexagon of order two, and black holes {\it Phys. Rev. D} {\bf 78} 124022.

\bibitem{PlanatCPT}
Planat M 2009 hree-qubit entangled embeddings of CPT and Dirac groups within E8 Weyl group  Preprint 09043691 [quant-ph]. 

\end{thebibliography}
\end{document}